# Proximity-induced supercurrent through topological insulator based nanowires for quantum computation studies

Biplab Bhattacharyya,[1,2] V. P. S. Awana,[1,2] T. D. Senguttuvan,[1,2] V. N. Ojha,[1,2] Sudhir Husale[1,2]*

[1] Academy of Scientific and Innovative Research (AcSIR), National Physical Laboratory, Council of Scientific and Industrial Research, Dr. K. S Krishnan Road, New Delhi-110012, India.

[2] National Physical Laboratory, Council of Scientific and Industrial Research, Dr. K. S Krishnan Road, New Delhi-110012, India.

*E-mail: husalesc@nplindia.org

**Abstract**

Proximity-induced superconducting energy gap in the surface states of topological insulators has been predicted to host the much wanted Majorana fermions for fault-tolerant quantum computation. Recent theoretically proposed architectures for topological quantum computation via Majoranas are based on large networks of Kitaev's one-dimensional quantum wires, which pose a huge experimental challenge in terms of scalability of the current single nanowire based devices. Here, we address this problem by realizing robust superconductivity in junctions of fabricated topological insulator ($Bi_2Se_3$) nanowires proximity-coupled to conventional s-wave superconducting (W) electrodes. Milling technique possesses great potential in fabrication of any desired shapes and structures at nanoscale level, and therefore can be effectively utilized to scale-up the existing single nanowire based design into nanowire based network architectures. We demonstrate the dominant role of ballistic topological surface states in propagating the long-range proximity induced superconducting order with high $I_cR_N$ product in long $Bi_2Se_3$ junctions. Large upper critical magnetic fields exceeding the Chandrasekhar-Clogston limit suggests the existence of robust superconducting order with spin-triplet cooper pairing. An unconventional inverse dependence of $I_cR_N$ product on the width of the nanowire junction was also observed.

**Introduction**

The increasing demand for faster and more powerful computational techniques has motivated researchers throughout the globe to shift attention towards quantum computation (QC) studies. Some of the recent developments along this direction include quantum bit (qubit) realization through photon polarization, cavity quantum electrodynamics, nuclear magnetic resonance, magnetic flux based superconducting quantum interference devices (SQUID), etc.[1] Though, these experiments provide an exciting and optimistic roadmap towards QC, but are prone to computational errors arising from the imperfections or disorder or thermal/charge fluctuations in the quantum-mechanical system, environmental interactions, imprecise gate operations, etc. leading to de-coherence of qubits.[1] Therefore, the error protection needs to be implemented at the hardware level itself.[2] Theoretical predictions state that some topological phases of matter obey non-Abelian exchange statistics, where the non-commutative interchange of particles helps to implement error control in order to perform topological quantum computation (TQC).[2,3] Recently, Aasen et al[4] and Karzig et al[5] have listed certain basic milestones that are necessary to make TQC a reality. These are: 1) detection/engineering of non-Abelian anyons/quasi-particles and their validation as topological qubits, 2) processing of quantum information via braiding of these quasi-particles through fabricated junctions and networks using gate voltages, and 3) determining the overall result of the braiding mechanism. Majorana fermions (fermions that are their own anti-particles, i.e. unlike Dirac fermions, Majorana fermions have equal creation and annihilation operators, $\gamma^{\dagger} = \gamma$)[2,6] are predicted to occur as zero-energy excitations (Majorana zero modes, MZMs) in vortices of two-dimensional (2D) spinless p+ip superconductors; fractional quantum Hall effect (FQHE, $\nu = 5/2$ state); and Kitaev's model of 1D p-wave superconducting nanowire in topological phase.[2,3] Interestingly, these MZMs are expected to exhibit non-Abelian braiding statistics.[2,3] Owing to the difficulty in implementation of p-wave superconductors, elusive nature of $\nu = 5/2$ quantum Hall state and the recent advances in the nanofabrication techniques; the interfaces of semiconductors and ordinary s-wave superconductors have gained widespread popularity for hosting MZMs.[6,7]

Experimental observation of MZMs (zero-bias peak (ZBP) in differential conductance measurements) in 1D semiconducting nanowires of strong spin-orbit coupling (SOC) systems such as InAs[7] and InSb[6] in proximity to an s-wave superconductor was a major breakthrough in the field of TQC. But, research has shown that disorder related effects such as Kondo effect, weak-antilocalization (WAL) and Andreev bound levels close to zero energy also give rise to a ZBP in Majorana quantum wires, which may interfere with the Majorana bound zero energy state and thus account for the non-quantized ZBP ($\neq 2\,e^2/h$) observed in these experiments.[6,8] Although, the development of a ballistic Majorana nanowire device of InSb grown by Au-catalysed vapour-liquid-solid (VLS) mechanism[8] overrule the disorder related effects, but the practical Majorana-based future TQC devices will mostly rely upon nanofabrication techniques like lithography and milling, where disorder will play a crucial role in hampering the topological phase of such semiconductor-based 1D Kitaev chains.

Recently, topological insulators (TIs) have emerged as one of the most promising candidate materials for QC due to the existence of topologically protected surface conduction that is immune to non-magnetic local perturbations/disorder in the material.[9] These topological surface states (TSS) show high robustness due to the strong spin-momentum locking and time reversal symmetry (TRS), which prevents backscattering events on the surface and provides way for dissipationless charge and spin currents.[9] Till date, many observations of these exotic surface states (SS) in TI materials like $Bi_2Se_3$, $Bi_2Te_3$, $Sb_2Te_3$, SnTe, $Bi_{2-x}Sb_xTe_{3-y}Se_y$, etc. have been demonstrated.[9] Most of these experiments were performed on TI-based nanowires, nanoribbons, nanoflakes, nanoplates or nanotubes, where the TSS conductance contribution is significantly enhanced due to large surface-to-volume ratio. Previously, in a groundbreaking work by Fu and Kane,[10] it was theoretically shown that the non-Abelian MZMs are also supported at the interfaces of 3D TIs in proximity to an s-wave superconductor. Recent experimental detection of MZMs in the magnetic vortex of a 3D TI ($Bi_2Te_3$)/superconductor heterostructure[11] has triggered the notion that TI- nanostructures can make as efficient hardware material for fault-tolerant TQC. Unlike semiconducting materials, where the spinless condition for electrons is met by tuning the chemical potential inside the small Zeeman gap induced via external magnetic field,[6,7] TIs intrinsically possess spin-non-degenerate SS even in the absence of external magnetic field.[9] Inducing superconductivity in these SS will not only host the Majorana fermions, but also the resulting superconducting order will be robust to the non-magnetic perturbations or imperfections in the system. Thus, the topologically protected robust TSS provides a more favourable and efficient platform to realize MZMs and perform TQC in comparison to other high SOC semiconductors.

Another interesting approach towards the realization of MZMs in topological superconductors (TSC) is the fractional Josephson effect, which has a 4π-periodic current-phase relationship (CPR) rather than the normal 2π-periodicity.[3,12] This effect was first predicted by Kitaev for an ideal 1D spinless p-wave superconductor[3] and later by Fu and Kane for superconductor/quantum-spin-Hall-insulator/superconductor[12] based Josephson junctions. Significant experimental work has been performed along this direction in the last few years, where Josephson supercurrent was realized in exfoliated flakes of $Bi_2Se_3$[13], $Bi_2Te_3$[14], $Bi_{1.5}Sb_{0.5}Te_{1.7}Se_{1.3}$[15] and strained bulk HgTe[16]. Also, an anomalous CPR and unconventional Josephson effect was observed in the weak links of exfoliated $Bi_2Se_3$ flakes[13] and strained HgTe layers[17]. All these experiments have used TI-based flakes or thin films as the weak link, whereas practical TQC devices will be based on 1D nanowires of proximity-induced TSC. Till date, experiments demonstrating the existence of proximity-induced superconductivity in nanowires of TIs are lacking. Therefore, realizing supercurrent through fabricated Josephson junctions of superconductor-TI nanowire-superconductor will be a major step towards the detection of MZMs and realistic TQC hardware.

In this work, we demonstrate for the first time, the existence of long-range proximity-induced supercurrent through fabricated nanodevices of tungsten (W)-$Bi_2Se_3$ nanowire-W junctions, useful for quantum computing studies. We incorporate the focused-ion-beam (FIB) milling method to fabricate nanowires from mechanically exfoliated thin flakes of $Bi_2Se_3$.[18-21] Various Majorana braiding architectures such as 1D wire networks forming a T-junction[2], hexon architecture[5], tri-junction geometry[4], etc. have been proposed in the recent past. Ion-milling technique provides an excellent way to experimentally realize such architectures in comparison to other nanowire synthesis routes like chemical vapour deposition (CVD), VLS or electro-deposition, where fabricating such braiding networks is not possible. The observation of proximity-induced superconductivity in long $Bi_2Se_3$ junction lengths with high $I_cR_N$ products indicates the preferred coupling of proximity effect with the ballistic TSS channels, which is an essential requirement for a TSC to host MZMs. An unconventional inverse dependence of the $I_cR_N$ product on the width of the nanowire, suggests the possibility of 1D-confined quantum states at the superconductor-TI (S-TI) interface. Also, the estimated large upper critical field $B_{c_2}(0)$ (greater than the Chandrasekhar-Clogston limit) demonstrates the existence of proximity-induced robust superconducting order in TIs with a possible spin-triplet cooper pairing that is crucial for hosting MZMs.

**Results**

False colored FESEM images of the weak links of $Bi_2Se_3$ nanowires (devices N4 and N5) with four-probe measurement geometry are depicted in insets of Fig. 1a. The estimated length (i.e. junction length) and width of the weak links ($Bi_2Se_3$ nanowires) is about 368 nm and 226 nm for N4, and 286 nm and 121 nm for N5, respectively. Figure 1a shows the resistance vs. temperature (RT) characteristics (10 K to 2 K) of the samples N4 and N5. A small resistance increase of 47.08 Ω (N4) and 30.62 Ω (N5) is observed while cooling the system from room temperature to 15 K, followed by a sudden superconducting transition ($T_c^{onset}$) at 5.357 K (N4) and 5.706 K (N5). This is consistent with the previously reported $T_c$ values of ~ 5 K for FIB-deposited W.[22,23] The transition width, i.e. $\Delta T_c = T_c^{onset} - T_c^{zero}$, is smaller for N5 (0.65 K) than N4 (2.81 K), which is evident from the shorter junction length of N5. Figure 1b depicts the perpendicular magnetoresistance (MR) plot at 2 K for sample N5. The MR characteristics include both superconductivity and TSS properties, with zero resistance around zero B-field, above which MR increases with B-field. MR is symmetric with applied B-field direction. MR curve has been divided into three regimes with increasing B-field. Regime 1 represents the state of complete superconductivity in the nanowire due to proximity effect. Resistance starts to increase linearly with B-field in regime 2. The linear MR has been reported in various TI materials and is attributed to the linear Dirac energy spectrum of the TSS.[9,18] Also, it is well known that TIs exhibit Shubnikov-de Haas oscillations at higher perpendicular B-fields due to the presence of metallic TSS.[9,18] Similar oscillatory features can be observed in regime 3 of the MR curve. Thus, the MR at 2 K reveals the characteristics of TSC arising from the proximity-induced superconductivity in TSS of the $Bi_2Se_3$ nanowire.

Figure 2 depicts the broadening in RT curves with increasing perpendicular magnetic field (B) up to 14 T. The shifting of $T_c^{onset}$ towards lower temperatures with increasing B-field is clearly visible for the samples. Zero resistance state is lost above 4 T for both N4 and N5 in the available temperature range till 2 K. Interestingly, for device N4, we observe a small resistance increase (kink-type feature) at 4.9 K (0 T). This resistance kink disappears above 0.5 T (upper inset in Fig. 2a), which suggests its dependence on the applied perpendicular B-field. In order to calculate the upper critical field ($B_{c_2}$), we define $T_c$ at 90 % of the resistance transition, i.e. $R = 90\% R_N$, where $R_N$ stands for the normal state resistance with values around 445 Ω for N4 and 556 Ω for N5. We incorporate the three standard approaches to calculate $B_{c_2}$: i) normal state paramagnetism leading to a Pauli limited transition field due to the competing Zeeman energy of external magnetic field and cooper pair condensation energy, given by the Chandrasekhar–Clogston limit,[24]

$$B_{c_2}(0)^2 N(E_F) = \frac{1}{2}\Delta(0)^2 N(E_F), \qquad (1)$$

where $N(E_F)$ is the density of states at Fermi level and $\Delta(0)$ is the superconducting energy gap at T = 0. If we assume weakly coupled Bardeen-Cooper-Schrieffer (BCS) limit, $\Delta(0) = 1.76 k_B T_c$ ,[25-27] then Eq. (1) leads to the Pauli paramagnetic limit[24,28] given by $B_{c_2}^P(0) = 1.84\, T_c = 9.856$ T (10.499 T) for N4 (N5) at 0K; ii) $B_{c_2}(0)$ for a type-II superconductor in dirty limit is given by the single-band Werthamer-Helfand-Hohenberg (WHH) theory,[29]

$$B_{c_2}(0) = -0.693\, T_c \left(\frac{dB_{c_2}}{dT}\right)_{T=T_c}, \qquad (2)$$

where $B_{c_2}(0)$ is estimated to be 19.75 T (11.18 T) for N4 (N5); iii) generalised Ginzburg-Landau (GL) equation, given by:[30]

$$B_{c_2}(T) = B_{c_2}(0) \left[\frac{1-t^2}{1+t^2}\right], \qquad (3)$$

where $B_{c_2}(0)$ is the upper critical field at 0 K and t = reduced temperature ($T/T_c$). Lower inset in Figs. 2a and 2b shows the extrapolated GL equation fitting (solid line) to the experimental data (spheres) giving a $B_{c_2}(0)$ value of 28.5 T (16.14 T) for N4 (N5). Since GL theory provides the highest $B_{c_2}(0)$ values, the superconductivity in these samples can be described under the GL theory regime. Such high values of $B_{c_2}(0)$ reflects the robustness of this proximity-induced topological superconducting state towards high magnetic fields, which is necessary to sustain and manipulate

MZMs via magnetic field. The superconducting coherence length $\xi(0)$ can be estimated from the $B_{c_2}(0)$ value using the GL formula[23]

$$\xi(0) = \sqrt{\frac{\phi_o}{2\pi B_{c_2}(0)}}, \qquad (4)$$

where $\phi_o$ is the flux quantum related to a cooper pair (h/2e). Equation (4) gives a value of 3.4 nm and 4.52 nm for N4 and N5, respectively. The value is consistent with the previously reported values of coherence lengths in W-induced proximity superconductors.[23] Proximity-induced gap along the length of a nanowire in a superconductor-normal metal-superconductor (SNS) system depends on the barrier coherence length ($L_T$), i.e. the decay length of proximity-induced order parameter in N, given by:[22,31]

$$\Delta(x) = \Delta_L \frac{\cosh(x/L_T)}{\cosh(L/L_T)}, \qquad (5)$$

where length of the nanowire is 2L (centre at x=0) and $\Delta_L$ is the superconducting gap at the N-S interface. This equation clearly suggests that the proximity-induced superconductivity in the nanowire will weaken with increasing length. At the centre of the nanowire, $\Delta(x=0) \sim [\cosh(L/L_T)]^{-1}$ indicates that smaller values of $L_T$ will destroy the superconducting state. In case of clean and highly transparent N-S interfaces, long range proximity effects up to 1 μm can be observed.[22,32,33] Previously, formation of spin-triplet cooper pairs in ferromagnetic Co nanowires was shown to possess long range proximity-induced superconductivity (~1 μm).[32] Similar kind of phenomenon is possible in TIs, where the spin-polarized surface current can lead to a spin-triplet superconducting pairing of the SS. Theoretically, unconventional superconducting order is predicted to exist in the TSS though proximity effects, which has the potential to host MZMs.[10] Kasumov et al.[33] reported the first evidence of an anomalous proximity effect in a TI material BiSb, where the critical current ($I_c$) was shown to increase with increasing junction length, unlike Eq. (5). The reason for this anomaly was attributed to the very low effective mass of electrons in such materials (now called TIs), that accounts for high de-Broglie wavelengths comparable to the junction lengths.[33] Also, recent reports have shown the existence of long-range proximity-induced superconductivity in TI-based weak links, even for junction lengths much longer than the barrier coherence length ($L_T$) in diffusive channels and electron phase breaking length ($L_\phi$).[14,26] This inconsistency with the existing theory of superconducting proximity effect in normal metals can be explained by considering two different transport channels in TIs, ballistic TSS and diffusive bulk. Although, it is difficult to separate the diffusive bulk transport channels, but it was shown that the proximity-induced superconducting order prefers the ballistic TSS channel.[26] A similar kind of explanation is valid for the TI devices studied in this work. If a ballistic TSS channel is considered, then $L_T = \hbar v_F/2\pi k_B T \sim 304$ nm at 2K for $v_F \sim 5 \times 10^5$ ms$^{-1}$ (Fermi velocity in TSS for $Bi_2Se_3$)[18,26] is comparable to the junction length of the devices, where $\hbar$ = reduced Planck's constant and $k_B$ = Boltzmann constant = 8.62 x 10$^{-5}$ eV/K. This indicates towards the possibility of proximity-induced superconducting order parameter Δ(x) to survive throughout the long junction lengths due to the preferred coupling with the TSS, which is very crucial for hosting the MZMs that are predicted to occur when proximity effect induces superconductivity in the TSS.

In order to understand the supercurrent behaviour in our samples, we hereby present the current-voltage (IV) characteristics, i.e. voltage versus the applied current. Figure 3a depicts the IV curves (device N4) for temperatures ranging from 2 K to 5.8 K at B = 0. Supercurrent vanishes with increasing temperature and the curves follow typical linear ohmic behaviour above 4.6 K. At 2 K, $I_c$ is 1.641 µA, above which there appears a finite voltage, as the system is driven to a resistive state. The IV curves for N4 depict low voltage foot-like features (shown by pink arrow in Fig. 3a), which was first observed by Octavio, Skocpol and Tinkham[34] in tin-microbridges and was attributed to the non-equilibrium quasiparticles created by the breaking of cooper pairs during the Josephson cycle. Figure 3b shows the IV curves for device N5 at different temperatures. A very sharp transition from superconducting to resistive regime can be observed for N5 at lower temperatures, which is obvious from the sharp superconducting RT transition in this sample as shown above in Fig. 2b. For temperatures above 3.4 K, foot-like transition starts to appear (shown by pink arrow in Fig. 3b). The $I_c R_N$ product ($I_c$ = critical current and $R_N$ = normal state resistance), which provides information about the superconducting gap (Δ), is equal to 0.73 mV for N4 at 2 K. Previously, it was reported that the maximum $I_c R_N$ product at low temperature in a SNS junction is πΔ/e, where e is the electronic charge.[25] If we assume that BCS relation is followed by the superconducting contacts in this case, then $\Delta(T = 0) = 1.76 k_B T_c$ provides a gap of 0.8127 meV for N4 at $T_c$ corresponding to 90% of resistance transition = 5.357 K ($k_B$ = Boltzmann constant = 8.62 x $10^{-5}$ eV/K). Thus, the expected maximum $I_c R_N$ product for N4 is 2.55 mV, which is about 3 times higher than the value at 2 K. A high $I_c R_N$ product of 4.6 mV is obtained for N5 at 2 K ($I_c$= 8.32 µA). The superconducting gap (Δ) of 0.866 meV is estimated for N5, and maximum $I_c R_N$ product of 2.72 mV, which is about 1.5 times smaller than the value at 2 K. The expected maximum $I_c R_N$ product at 0 K is slightly less for N4 due to the longer junction length than N5. Since N4 has longer junction length than both the ballistic TSS $L_T$ and diffusive bulk $L_T$ at 2 K, therefore, the $I_c R_N$ product for N4 is very less at 2 K. Whereas, N5 has a shorter junction length than the estimated ballistic TSS $L_T$ (~ 304 nm at 2 K) that leads to high $I_c R_N$ product at 2 K. This again hints towards the preferable coupling of proximity effect to the TSS channels, which is the basic need for the detection of Majorana fermions. An alternative explanation for the suppressed $I_c R_N$ product in N4 can be the insufficiently transparent S-TI interfaces due to the presence of tiny superconducting islands near the electrode edges from FIB-deposition of W electrodes. Previously, anomalous small $I_c R_N$ products were observed for TI flake or thin film based Josephson junctions.[13,14,16] Also, it is well known that this value is independent of the junction geometry.[27] But, Williams et al.[13] reported an inverse dependence of $I_c R_N$ with the width of the TI weak link. We observe a similar dependence, where $I_c R_N$ product (at 2 K) is larger for device N5

(width = 121 nm) than N4 (width = 226 nm). Since the minimum temperature accessible to us in this experiment is 2 K, therefore, we can assume a higher $I_c R_N$ product at further lower temperatures.

Upper insets in Figs. 3a and 3b depict the clean and dirty limit fits using the de Gennes $I_c$ equation for SNS junctions:[31]

$$I_c = \frac{\pi \Delta}{2eR_N} \frac{L/L_T}{\sinh(L/L_T)}, \quad (6)$$

where e is the electronic charge, L represents the junction length and $L_T$ represents the barrier coherence length. For clean or ballistic junctions $L_T = \hbar v_F / 2\pi k_B T$, and for dirty or diffusive junctions $L_T = \sqrt{\hbar D / 2\pi k_B T}$, where $\hbar$ = reduced Planck's constant, $v_F$ = Fermi velocity in N and D = diffusion coefficient.[14,15,25,31,35] Since, Eq. (6) is based on the GL theory; therefore ideally it should be valid only near $T_c$. But, reports have shown its validity even at very low temperatures.[35] For our case, we found the best $I_c$ vs. T fit to Eq. (6) for 2 K ≤ T ≤ 3.4 K. For N5, it can be seen that the experimental data diverges from the fits for T > 3.4 K. Due to lack of experimental data below 2 K, we are unable to distinguish whether the system is in clean or dirty limit from the fit, as both the limits fit to the data in 2 K ≤ T ≤ 3.4 K. As explained in the Methods section, the presence of small superconducting islands nearby the S-TI interface due to FIB-deposited W may introduce disorder in the system. Thus, the transport in our junctions can be assumed to be slightly in the dirty/diffusive regime, where bulk transport channels might interfere with the surface transport. In that case, the value of D from the dirty limit fit is estimated to be 0.0137 m²/s and 0.059 m²/s for N4 and N5, respectively. The values are in agreement with the previously reported D values in $Bi_2Se_3$ material.[13] In a dirty/diffusive SNS junction, the Usadel equation (which gives information about the quantum transport in a dirty/diffusive superconductor, i.e. where junction length is longer than the electron mean free path)[14,15,25] can be written in terms of junction length as:[36]

$$I_c = \frac{64\pi k_B T}{eR_N} \sum_{n=0}^{\infty} \left(\frac{L}{L_{\omega_n}}\right) \frac{\Delta^2 e^{-L/L_{\omega_n}}}{\left[\omega_n + \Omega_n + \sqrt{2(\Omega^2 + \omega_n \Omega_n)}\right]^2}, \quad (7)$$

where Matsubara frequency for fermions, $\omega_n = (2n+1)\pi k_B T$ is the set of discrete imaginary frequencies of all the degenerate excited states present in the system at finite temperature. $L_{\omega_n} = \sqrt{\hbar D / 2\omega_n}$ is the barrier coherence length at different Matsubara frequencies with $L_{\omega_0} = L_T$. $\Omega_n = \sqrt{\Delta^2 + \omega_n^2}$ is energy related to the superconducting gap and thermally excited states at finite temperature. Dubos et al.[36] performed the numerical analysis of Eq. (7) and came up with an approximate solution for low temperature regime, given by:

$$I_c = \frac{a\hbar D}{eL^2 R_N} \left[1 - b \exp\left(\frac{-a\hbar D}{3.2 \, L^2 k_B T}\right)\right], \quad (8)$$

where a and b are unitless coefficients. Lower insets in Figs. 3a and 3b shows the fit to Eq. (8) for device N4 and N5, respectively. The fit for N5 deviates from experimental data when close to $T_c$ as

Eq. (8) is valid in the low temperature regime. Transport in diffusive conductors is governed by a characteristic energy scale, $E_c = \hbar D/L^2$, called the Thouless energy, which provides the diffusion rate through the system.[16,36,37] For our case, $E_c$ is estimated to be 0.0667 meV and 0.476 meV for N4 and N5, respectively. Since $E_c < \Delta$ ($\Delta$ = 0.8127 meV for N4 and 0.866 meV for N5) for both the devices, therefore, the systems are in the long junction limit,[36,37] where diffusive $L_T$ is smaller than the junction length of the devices. The existence of long-range superconducting proximity effect even with small diffusive $L_T$, clearly indicates the co-existence of two different types of transport channels (bulk and TSS) in the $Bi_2Se_3$ junction. As discussed earlier, TSS is expected to carry this long-range proximity-induced superconducting order parameter through the long TI junction.

Figure 3c depicts the hysteresis present in the IV characteristics of device N5 at 2 K. The critical current observed while switching from resistive to supercurrent regime (downward current sweep), i.e. the retrapping current $I_r$, is slightly less than the critical current ($I_c$) for supercurrent to resistive state transition (upward current sweep). Upper inset in Fig. 3c shows the variation in $I_r$ and $I_c$ with temperature, where small hysteretic behaviour disappears completely above 3.4 K. According to the RCSJ (resistively and capacitively shunted junction) model,[25,27] a Josephson junction can be described by resistance and capacitance connected in parallel to each other. In our case, the effective capacitance ($C_{eff}$) of the lateral junction can be written as

$$C_{eff} = \frac{\varepsilon_0 K_1 t_{sc} W_{sc}}{L} + \frac{\varepsilon_0 K_2 t_{nw} W_{nw}}{L}, \qquad (9)$$

where $\varepsilon_0$ is the vacuum permittivity, $K_1$ and $K_2$ are the dielectric constants of vacuum and $Bi_2Se_3$, $t_{sc,nw}$ and $W_{sc,nw}$ are the thickness and width of superconducting electrode and $Bi_2Se_3$ nanowire, respectively. For device N5, $C_{eff}$ is estimated to be 95.83 aF. Thus, the Stewart-Mc Cumber parameter ($\beta_c$) is estimated to be 0.747 at 2 K from Eq. (10).[16,25,27]

$$\beta_c = \left(\frac{2\pi}{\phi_o}\right) I_c R_N^2 C_{eff} \qquad (10)$$

Usually, hysteresis is observed in under-damped junctions ($\beta_c \gg 1$).[16,25,27] But in this case the junction is almost critically or intermediately damped ($\beta_c \approx 1$); therefore, the width of hysteresis loop is quite small. Since large hysteresis is related to large values of shunt resistance and capacitance that ultimately leads to decrease of $I_c$ (and superconductivity), therefore, presence of small hysteresis in superconducting TI junctions (similar to N5) provides favorable platform to host MZMs. Previously, Courtois et al.[37] demonstrated that the hysteresis in SNS junctions is a direct consequence of electron heating in the normal metal during the supercurrent to resistive regime transition. According to the Skocpol, Beasley and Tinkham (SBT) hotspot model,[38] the Joule heating in the weak link may induce an increase in the local temperature, where $\sqrt{1 - T/T_c}$ dependence of $I_r$ is predicted with temperature.[38,39] Upper inset in Fig. 3d depicts the $(1 - T/T_c)^{0.33}$ dependence of $I_r$ on T, which suggests the possibility of self-heating induced hysteresis for N5. No hysteretic IV behaviour was observed for device N4 in the available temperature range. Figure 3d shows the $(1 - T/T_c)^n$ dependence of $I_c$ on T for device N5. For clean superconductors, n=1.5 represents the standard GL model, where $I_c$ is dependent on the cooper pair breaking mechanism in the weak links.[39] In case of dirty superconductors, n exceeds 1.5. Since the GL theory is valid near $T_c$, therefore, we divide the data into two regimes, T < 3.5 K and T > 3.5 K. For T < 3.5 K, n equals 0.4, which does not follow the GL value. Near $T_c$ (T > 3.5 K), n=1.6, which corresponds to the GL value in dirty limit. It must be noted that for all the graphs (including upper inset) in Fig. 3d, $T_c$ was taken at maximum value of dR/dT. Lower inset in Fig. 3d shows the decrease in $I_c$ with increasing B-field at 2 K. Supercurrent is strongly retained at 1 T, which is suggestive of the high critical field ($B_c$) of the system. Previous reports used small external B-field perpendicular to spin-orbit B-field to open a Zeeman gap in semiconducting nanowires to detect MZMs.[6-8] Since this is essential for a Majorana wire to be in topological phase,[3,8] therefore, presence of strong supercurrent even at high B-field of 1 T suggests the efficiency of TI-based nanowires for hosting MZMs. Interestingly, we also observed phase slip events in device N5 up to 3.2 K. Previously, coherent quantum phase slip (CQPS) was demonstrated in narrow segments of disordered superconductor, where local inhomogeneities can give rise to phase slip centres (PSC).[40] The phase of the superconducting order parameter shifts by different rates on both side of PSC, and when the order parameter becomes zero at PSC, the phase difference slips by $2\pi$.[27] This manifests itself as regular voltage steps in the IV curve, once the current is increased above $I_c$. Lower inset in Fig. 3c highlights the voltage step due to phase slip. The FIB-induced disorder can account for the observed event in TI nanowire. Since phase slip event is exactly dual to the Josephson effect, therefore, phase slip mechanism can be used to realize the quantum current standard.[39,40] Although, the detailed study of phase slip events in TIs is not the prime motive of this work, but their observation in FIB-fabricated $Bi_2Se_3$ nanowire triggers many future experiments intended to complete the quantum metrological triangle, i.e. resistance, voltage and current standard via TI-based quantum wires.

**Discussion**

Realization of supercurrent in FIB-fabricated nanowires of TIs is a significant step towards the hosting of MZMs for TQC. It is predicted that in the topological phase of Kitaev's superconducting quantum wire, Majorana fermions from the neighbouring lattice sites will form bound pairs, leaving two unpaired Majorana fermions at the ends of the nanowire.[3] This model of 1D p-wave TSC nanowire provides more efficiency in locating the MZMs than other 2D system based approaches like FQHE (ν = 5/2 state) or magnetic vortices, as in 2D system they can appear anywhere in the sample and at the same time could be masked by other low-energy 2D quasi-particle excitations, which increases the difficulty level of detecting MZMs. A major challenge to build a practical Majorana-based quantum computer is the scalability of these laboratory level hardware designs. Recent detection of Majorana fermions were performed in systems where scaling of the device geometry is very difficult. Along this line, the experts have recently proposed scalable designs for the realization of TQC with MZMs.[2,4,5] For the experimental realization of these designs, the fabrication of superconductor-TI nanowire junctions is of utmost priority. Presently, there is no single synthesis technique that can scale up the simple nanowire based devices of TI or other semiconducting material into complicated network geometries. To achieve high device scalability, one needs to grow or assemble 1D nanowires at specific locations, which surely is a very crucial step towards future quantum computing. Recently, we have shown and pioneered TI-based nanowire device fabrication by using milling method,[18-21] which has device fabrication potential for measuring two or more 1D nanowire based devices. We have already demonstrated the robust TSS properties in FIB fabricated nanowires of $Bi_2Se_3$ through ½-shifted Shubnikov-de Haas (SdH)[18] and Aharonov-Bohm (AB) oscillations[19] in our previous reports, which further encourages the use of ion-milling for building TQC architectures. It is important to note that the alignment of two or four nanowires and making device geometry for manipulation of MZMs is a very difficult task, but our proposed approach will solve this engineering problem and will give an access to perform experiments by braiding these quasi-particles, which is a milestone step that has never been accomplished before. Further, the direct FIB deposition or e-beam based masked design of superconducting contacts on these milled nanostructures will enable the proximity-induced superconductivity, a perfect hardware for the creation, detection and braiding of MZMs. Figure 4 depicts the fabrication approach proposed by us for building the TQC hardware. Figure 4a shows the recently proposed TQC architectures like one-sided hexon,[5] tri-junction geometry[4] and hexagon[10]. False colored FESEM images in Fig. 4b show the patterning of the recently proposed TQC architectures (shown in Fig. 4a) through FIB milling on exfoliated $Bi_2Se_3$ flakes. Figure 4b clearly demonstrates the superiority of milling technique, in arrangement of nanowires in a prescribed manner, over other nanowire synthesis methods. It must be noted that, higher quality thin films of TI via CVD or molecular-beam-epitaxy (MBE) can also be used for ion-milling of nanostructures. Figure 4c depicts the schematic of twelve FIB-milled TI nanowires (false colored FESEM image from Fig. 4b) with superconducting electrodes to host twenty-four bound pairs of Majorana fermions, which will serve as twelve qubits (2n MZMs = n-qubits). Braiding and reading the states of qubits with the proposed twelve nanowire array has the potential to compute much more data at a faster rate than the present day classical computer with millions of transistors. Realization of supercurrent in single nanowire of $Bi_2Se_3$ indicates that such FIB-fabricated architectures coupled with superconducting electrodes and proper gate voltages can be used to probe and braid MZMs.

Proximity-induced superconductivity in the TSS is predicted to lead to exotic phenomenon of Majorana fermions.[10] The Hamiltonian describing the proximity induced superconducting order parameter in TSS was earlier studied by Fu and Kane and can be written as[10]

$$\mathcal{H} = -iv\tau_z\sigma.\nabla - \mu\tau_z + \Delta_o(\tau_x \cos\varphi + \tau_y \sin\varphi), \qquad (11)$$

where the first term represents linear Dirac dispersion of TSS ($v$ = Dirac cone velocity, $\sigma$ = Pauli spin matrices), second term represents the chemical potential of the system ($\mu$ = chemical potential) and the third term represents the proximity-induced superconducting gap in the TSS ($\Delta = \Delta_o e^{i\varphi}$ denotes the superconducting energy gap with superconductor phase $\varphi$). $\tau_{x,y,z}$ is related to the particle-hole symmetry. Commutator $[\mathcal{T}, \mathcal{H}] = 0$ indicates that the overall Hamiltonian of the TI-superconductor heterostructure with proximity-induced superconductivity (Eq. (11)) follows TRS protection, where $\mathcal{T} = i\sigma_y K$ represents the time reversal operator with $K$ = complex conjugation. The spin-helicity in the TSS, $c_\pm = (c_{k\uparrow} \pm e^{-i\theta_k} c_{k\downarrow})/\sqrt{2}$, modifies the proximity-induced pairing term to[41]

$$H_\Delta = \sum_k \Delta\left(e^{i\theta_k} c^\dagger_{k,+} c^\dagger_{-k,+} + e^{-i\theta_k} c^\dagger_{k,-} c^\dagger_{-k,-} + H.c.\right), \qquad (12)$$

where $k$ = wave-vector, $c/c^\dagger$ = annihilation / creation operator, $\tan\theta_k = k_x/k_y$, $\uparrow\downarrow$ represents up and down spin states and $H.c.$ represents the Hermitian conjugation. Equation (12) represents an ideal spinless $p_x + ip_y$ superconductor. Although, the typical p-wave superconducting triplet pairing breaks TRS, but this Fu-Kane p-wave superconductor (proximity-induced superconductivity from s-wave superconductor into TSS) perfectly follows TRS. Thus, provides the required immunity and robustness to the system against small disorders and de-coherence related problems, which is essential for MZMs to perform error-free TQC. Though, direct measurements towards the detection of MZMs are out of scope for this study, but we have found experimental features that indicate towards the possibility of TSS propagated supercurrent in the $Bi_2Se_3$ nanowire junctions, which is a necessary condition for hosting MZMs. The high $I_cR_N$ product and long-range proximity effect even for junctions longer than the diffusive bulk $L_T$ suggests the dominant role of ballistic TSS channels (where $L_T$ is comparable to junction length) in our samples, which are immune to local perturbations/disorder in the system and thus can sustain the proximity-induced superconducting order over long junction lengths. The high $I_cR_N$ product, not following the conventional BCS weak-coupling limit of $\pi\Delta/e$, may be a result of the unconventional superconductivity induced in the TSS. Also, the observed unconventional inverse dependence of $I_cR_N$ product with the width of the nanowire was previously explained via quantum confinement along the 1D Majorana modes at the S-TI interface with quantized energy levels at multiples of $\varepsilon \propto hv_F\Delta/2\mu W$, where h = Planck's constant, $\Delta$ = superconducting gap, $\mu$ = chemical potential and W = width of the $Bi_2Se_3$ nanowire.[13] Under such conditions, the $I_cR_N$ product is governed by these confined Majorana modes, given by $\varepsilon/e$. Thus, inverse width dependence $I_cR_N \propto 1/W$ is observed for the TI-based junctions studied here. The Chandrasekhar-Clogston limit or Pauli paramagnetic limit is related to the breaking of spin-singlet cooper pairs as soon as the B-field crosses a critical value that leads to the aligning of electron spins along the B-field direction. Therefore, usually the superconductors have $B_{c_2}(0)$ value within this limit. Previously, the $B_{c_2}(0)$ values reported for FIB-deposited tungsten (W) nanowire (~ 9.5 T)[23] and Sr-doped $Bi_2Se_3$ crystals (~ 1.4 T)[30] were well within their Pauli paramagnetic limits. But in our case, the $B_{c_2}(0)$ value (28.5 T for N4 and 16.14 T for N5) is much greater than the Pauli paramagnetic limit for the respective devices. The high $B_{c_2}$ values indicate towards the possible existence of spin-triplet cooper pairing in the TI-superconductor system used in this study. Since, spin alignment along the field direction for a triplet pair is quite easy, therefore, the B-field required to break the triplet pair is expected to be very high. Thus, the high $B_{c_2}(0)$ values reported in this study not only demonstrates that robust superconductivity can be induced in TIs via proximity effect, but also suggests the possibility of spin-triplet cooper pairing in these proximity-coupled TI systems that is the primary pre-requisite for hosting MZMs. In general, spin-triplet cooper pairing is expected to exist in presence of spatial inversion symmetry in the material.[42] But, Frigeri et al.[43] have shown that triplet pairing is not completely absent under broken inversion symmetry. Rather, there may exist some spin-triplet pairs, which are not affected by the broken inversion symmetry. Another report found that due to broken 3D inversion symmetry at the 2D superconductor surface, a mixture of both spin-singlet and spin-triplet cooper pairing is present in the cooper pair wavefunction.[44] Additionally, previous theoretical reports demonstrated that in the absence of inversion symmetry, the Pauli paramagnetic limit is strongly suppressed,[43] which may also lead to $B_{c_2}(0)$ value exceeding the Pauli paramagnetic limit even for the spin-singlet cooper pairing. This triggers the need of further experimental work along this direction to correctly identify and understand the origin of superconducting symmetry (s-wave or p-wave or mixture of both) in TIs and other materials. The FIB-deposited W material also contains carbon and gallium impurities,[23] which alters the superconducting properties. To the best of our knowledge, only few reports have estimated the $B_{c_2}$ values, for e.g. Sun et al.[45] estimated $B_{c_2}$ value larger than 8 T at 2 K. Through our resistance-temperature-magnetic field (RTH) experiments, we have also observed $B_{c_2}$ more than 8 T. Moreover, we carried out RTH experiments up-to 14 T, which has been performed for the first time on FIB-deposited W material and estimated $B_{c_2}(0)$. Note that the reported $B_{c_2}(0)$ values are estimated using 90 % resistance transition value and hence, cannot be compared with the $B_{c_2}(0)$ value of parent pure W material. Here, we believe that more accurate values of $B_{c_2}(0)$ can be obtained via experiments carried out at ultra-low temperatures (mK) using high magnetic fields with different compositions of W, C and Ga impurities.

Another significant observation is the resistance kink in the RT characteristic of device N4 (upper inset in Fig. 2a). Previously, a similar kind of observation was attributed to the two different superconducting transitions, one from the W electrodes and other from the proximity-induced superconducting gap in the middle of nanowire.[22] But, such an intermediate re-entrant behaviour has also been shown to occur in the superconducting system with vortex states leading to a flux compression and paramagnetic Meissner effect (PME).[46] Though, the magnetization measurements showing PME are not performed in this work, but recent theoretical studies have shown the existence of odd frequency spin-triplet s-wave superconductivity in nanowire coupled to a topological superconductor, where odd frequency cooper pairs co-exist with Majorana fermions and give rise to PME.[47] In case of TIs proximity coupled to spin-singlet s-wave superconductor, it was theoretically shown that the odd frequency cooper pairs exist even in the absence of B-field and are robust against disorder.[48] Since, TI surface hosts spin-polarized current and W electrode has spin-singlet cooper pairs, therefore, spin-flip processes at the interface of $Bi_2Se_3$ nanowire and W electrode may also account for a small resistance increase leading to this kink-type feature. The accurate explanation for the origin of this feature is not possible in this work due to lack of experiments along this direction, but future magnetization experiments on similar kind of TI nanowire based system will shed more light on the exotic properties of odd frequency cooper pairs and Majorana fermions.

**Conclusion**

To conclude, we have studied the proximity-induced superconductivity in FIB-fabricated $Bi_2Se_3$ nanowire junctions with W electrodes. A sharp superconducting transition was observed for shorter junction length. The high $B_{c_2}$ values exceeding the Chandrasekhar–Clogston limit demonstrates the robust superconducting properties of proximity-coupled TIs with possible spin-triplet pairing of cooper pairs. The presence of long-range proximity effect with high $I_cR_N$ product in the junctions suggests the dominant role of ballistic TSS in propagating the supercurrent, even in the presence of diffusive bulk transport channels with barrier coherence length much smaller than the junction length of the devices. Also, the unconventional inverse dependence of $I_cR_N$ product on the width of the nanowire indicates the presence of quantum confined transport channel at the S-TI interface, which was previously predicted to arise from 1D Majorana modes at the interface. MR data also confirms the co-existence of both superconducting and TSS properties in the $Bi_2Se_3$ junctions, which is essential for hosting robust MZMs. Overall, our work on the fabrication of nanowires of TI, demonstration of the robust nature of 2D TSS transport[18,19] and supercurrent through TI-junction have huge potential to scale-up these nanodevices in any proposed geometry for building the fault-tolerant quantum computer and hence, stands as a breakthrough at the level of fundamental device physics and nano-engineering. The next step would be to observe the Majorana particles in similar kind of fabricated superconductor-TI nanowire junctions. Manipulation of non-Abelian MZMs are useful for encoding the quantum information, but movement or braiding of these particles has not been achieved yet, which is an essential ingredient for routine quantum computing operation. Supercurrent studies on nanowires of topological insulator and their synthesis reported by us establish a solid platform towards the study of superconductivity through topological surface states and the detection of Majorana fermions to perform braiding experiments. We believe that similar future experimental efforts and results will further enhance the expertise in qubit creation, detection and measurement capabilities for TQC.

**Methods**

**Device fabrication and characterization.** We have incorporated the FIB-based $Ga^+$ ion milling technique to fabricate the $Bi_2Se_3$ nanowire from exfoliated thin flakes of $Bi_2Se_3$. The standard Scotch-tape method was used to perform micro-mechanical cleavage of bulk crystals of $Bi_2Se_3$. The exfoliated flakes obtained were deposited on $SiO_2$/Si substrates with pre-defined gold contacts, which were already cleaned chemically (via acetone, iso-propanol, methanol and de-ionized water) and with oxygen plasma for 10 min. The thin flakes were localized under optical microscope (Olympus) and field emission scanning electron microscope (FESEM, Zeiss Auriga). FIB milling using $Ga^+$ ion (Zeiss Auriga) was further used to shape the flake in the form of a nanowire. The FIB-based gas injection system (GIS) was used to deposit the metal electrodes of tungsten (W) on the fabricated nanowire. To prevent any damages to the $Bi_2Se_3$ nanowire, a very low ion-beam-current was used (~ 50 pA) to deposit the superconducting electrode (W). Superconducting properties of W were optimized (thickness and $T_c$ ~ 5 K) before deposition and the same parameters were used for fabricating the devices. The release of precursor gas, tungsten hexa-carbonyl ($W(CO)_6$), was checked by monitoring the chamber vacuum. The FIB chamber vacuum was set to 5 x $10^{-6}$ mbar before releasing the gas from needle nozzle and 2 x $10^{-5}$ mbar during the deposition of W. This FIB induced metal deposition technique was further used to connect the W electrodes to pre-defined gold contacts, which serve as the functioning electrodes for the four terminal electrical measurements in the physical property measurement system (PPMS). In accordance with previous reports, where the FIB-deposited W material was shown to diffuse into the nanowire length close to the electrode edges,[22] it is visible in the high magnification FESEM images (See Fig. S1 in Supplementary Materials) that some of the superconducting (W) material has diffused along the length of $Bi_2Se_3$ nanowire up to a maximum of 40 nm from W electrode edge on either side. This diffusion length was also verified using the high-resolution transmission electron microscope (HRTEM) characterization of the FIB-deposited W electrode on $Bi_2Se_3$ flake (See Figs. S3 and S4 in Supplementary Materials). Therefore, the estimated length is calculated by subtracting this diffusion length from the actual junction length.

**Data availability.** All experimental data required to evaluate and interpret the conclusions are present in the main manuscript or supplementary materials file. Additional data or information related to this paper may be requested from the corresponding author (*E-mail: husalesc@nplindia.org).

**Acknowledgments**

We acknowledge Alka Sharma for assisting in the device fabrication process. B.B. is thankful to the Senior Research Fellowship from Council of Scientific and Industrial Research (CSIR), India. S.H. acknowledges the research support from CSIR-National Physical Laboratory.

**Author Contributions**

S.H. supervised the study; S.H. and B.B. fabricated the nanowire junctions. V.P.S.A. performed the PPMS measurements. T.D.S. and V.N.O. provided the FIB accessories, materials and laboratory facilities. Data analysis and manuscript preparation was done by B.B. and S.H. All authors read and commented on the manuscript.

**Additional Information**

**Supplementary information** containing the higher magnification FESEM images, HRTEM characterization of W-$Bi_2Se_3$ interface, absence of FIB-induced Halo effect during electrode deposition and magnetoresistance (MR) at 10K for device N5 accompanies the paper on the Scientific Reports website.

**Competing interests:** The authors declare no competing financial and non-financial interests.

# Figures

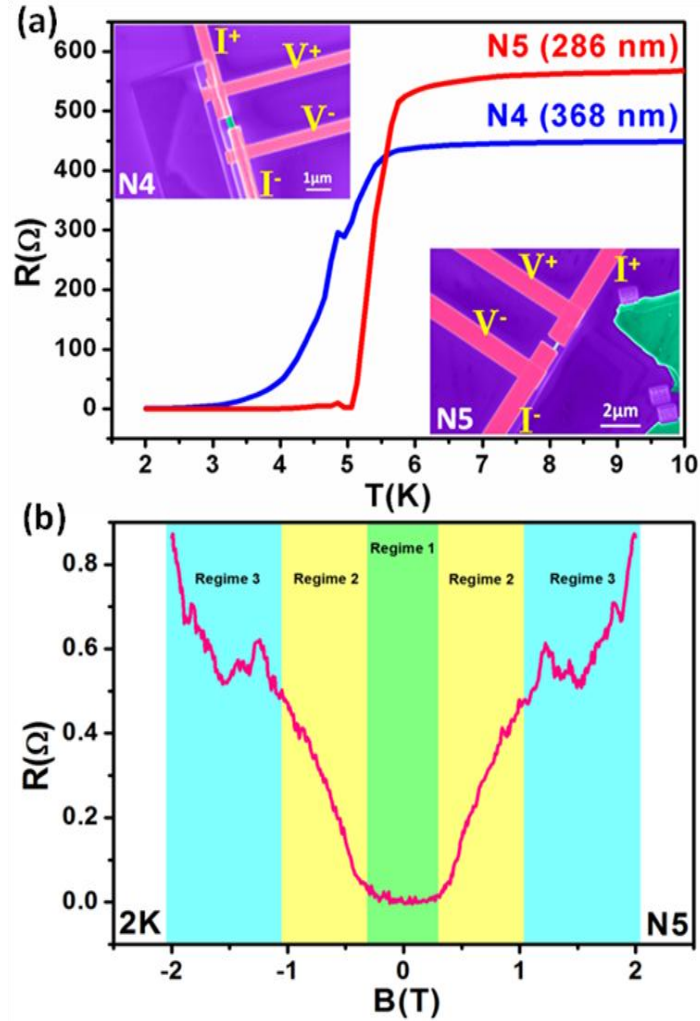

**Figure 1. Proximity-induced superconductivity in Bi$_2$Se$_3$ nanowire junctions.** (a) Cooling curves (10 K to 2 K) for devices N4 and N5 with junction lengths 368 nm and 286 nm, respectively, are shown. Both the devices show a superconducting transition at around 5 K. Insets depict the false-coloured FESEM images of the Bi$_2$Se$_3$ nanowire based tunnel junction devices (green colour) with superconducting W electrodes (pink colour) for four-probe measurements. (b) Topological superconducting characteristics in Bi$_2$Se$_3$ nanowire junction is also revealed by the MR data for device N5 at 2 K, which can be divided into three regimes. Regime 1 depicts the superconducting behaviour around zero-field. Regime 2 depicts the linear MR and regime 3 shows the MR oscillations at higher B-fields, which indicates the properties of metallic TSS.

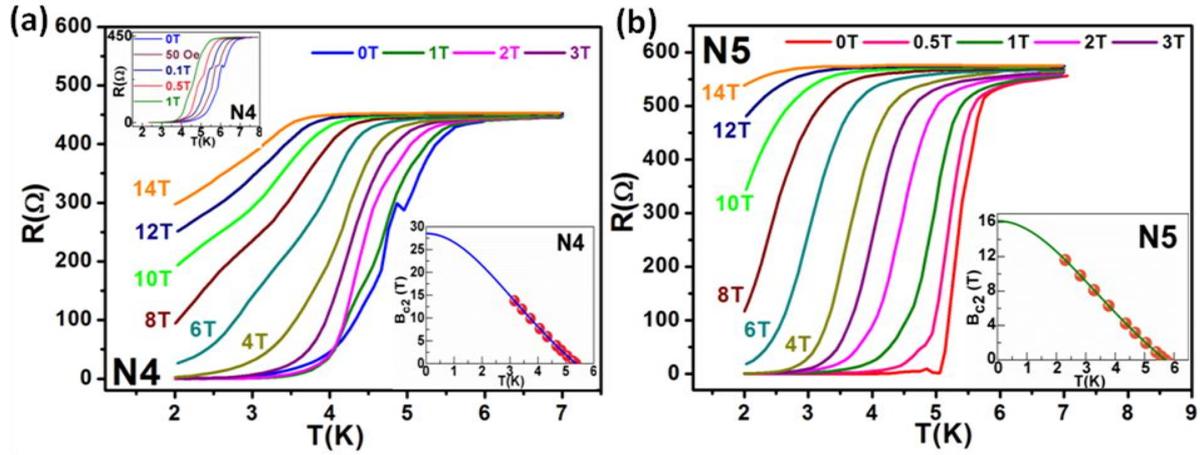

**Figure 2. Resistance versus temperature in presence of perpendicular magnetic field for critical B-field and supercurrent studies.** (a) RTH curves for N4 showing superconducting transition broadening with increase in B-field. Upper inset depicts the kink-type feature, which is observed in the resistance transition up to 0.5 T (curves shifted along x-axis by 0.3 K for clarity). Lower inset shows the extrapolated GL equation fitting to estimate the upper critical field $B_{c_2}(0)$. (b) RTH curves for N5 depict that the resistance drops rather sharply due to shorter channel length. Lower inset shows the GL fit for device N5.

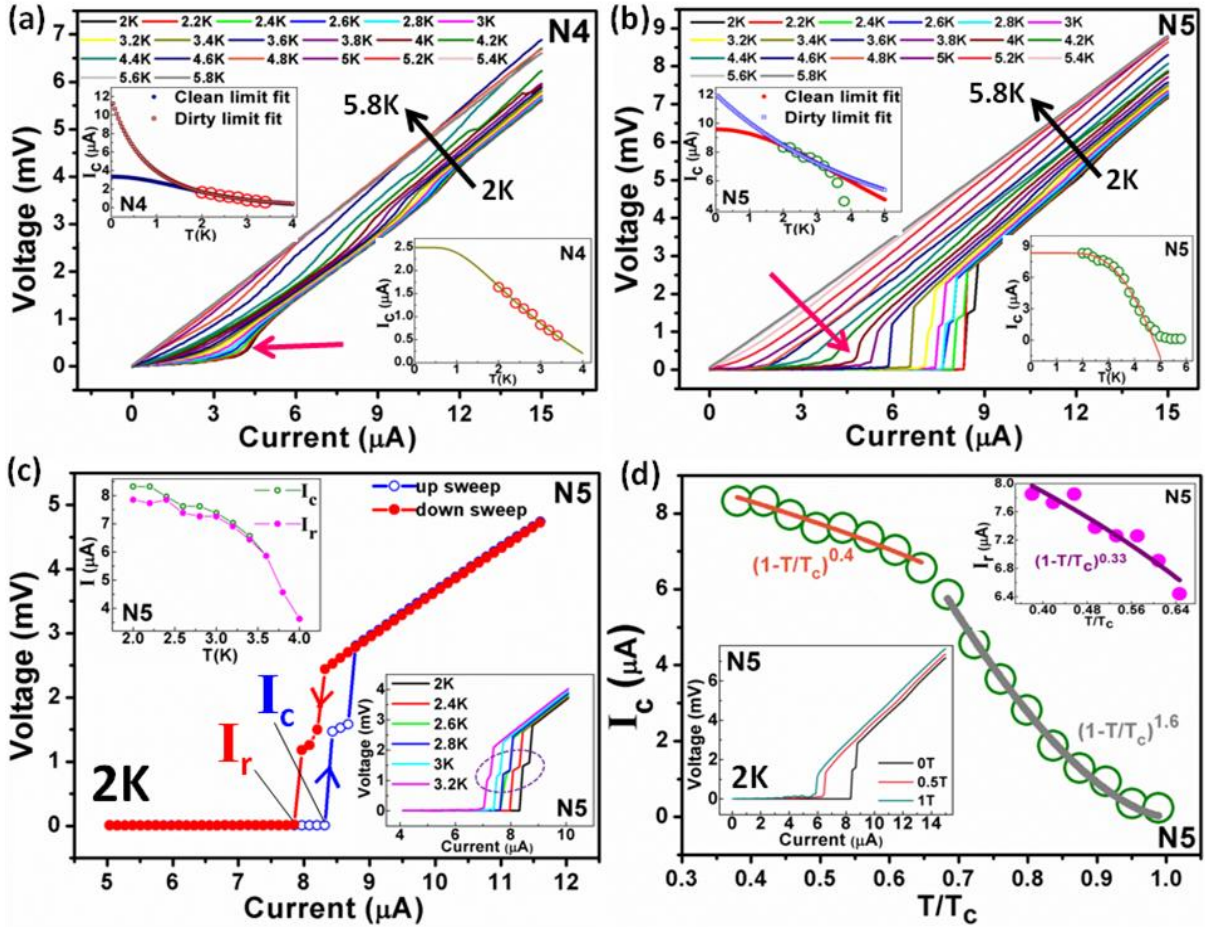

**Figure 3. Supercurrent analysis using current versus voltage (IV) characteristics. (a)** IV curves for device N4 show a clear foot-like (pink coloured arrow) superconducting transition. Upper inset depicts the extrapolated clean and dirty limit fits to Eq. (6). Lower inset shows the fit to Eq. (8), which gives information about the characteristic energy scale of the system. **(b)** IV curves for N5 depicting a sharp supercurrent transition till 3.6 K, above which foot-like transition (pink coloured arrow) starts to appear. At temperatures above $T_c$, a linear ohmic IV is observed. Upper and lower inset shows the fits to Eq. (6) and Eq. (8), respectively, where the fit starts to deviate from experimental data near $T_c$. **(c)** Hysteresis in device N5 at 2 K during up and down sweeps of current. Upper inset shows the small hysteresis persisting till 3.4 K, above which $I_r$ and $I_c$ merge. Lower inset highlights (dotted purple circle) the observed phase slip voltage step for 2 K < T < 3.2 K in device N5. **(d)** $I_c \sim (1 - T/T_c)^n$ fits at two temperature regimes, T < 3.4 K and T > 3.4 K, for N5. Upper inset shows similar fit for $I_r$. Lower inset depicts the decrease in $I_c$ with increasing B-field for N5.

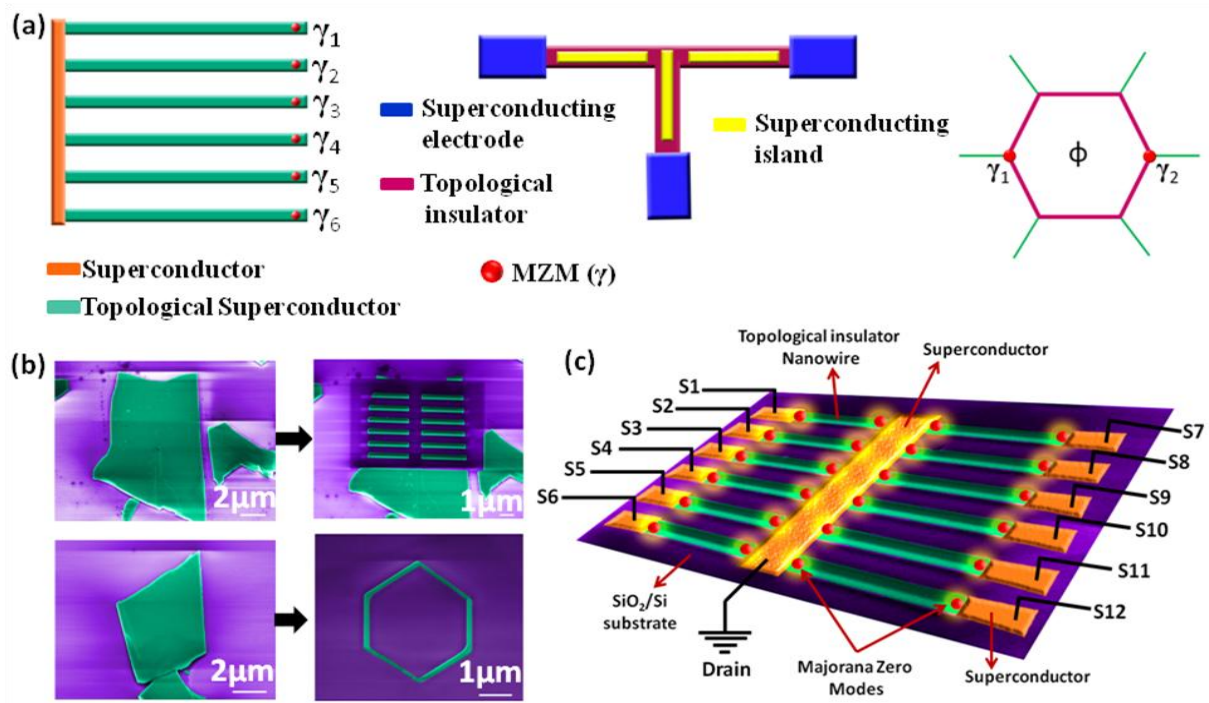

**Figure 4. Realization of recently proposed TQC architectures via FIB-milling.** (a) Recently proposed TQC architectures such as one-sided hexon, tri-junction geometry and hexagon for braiding MZMs. (b) False-coloured FESEM images of two-sets of six parallel nanowire arrays and hexagon fabricated by milling of the exfoliated $Bi_2Se_3$ flakes, which serve as the starting material for milling of TQC structures. The hexagon geometry was proposed to braid MZMs by modulating the phase ($\phi$) of the superconductor. (c) Schematic for hosting twenty-four Majorana bound states (twelve qubits) using FIB-fabricated TI ($Bi_2Se_3$) nanowire array shown in Fig. 4b. S denotes the source voltage. *Fig. 4a has been adapted from Ref. [5], [4] and [10], respectively.

# Supplementary Materials

**Proximity-induced supercurrent through topological insulator based nanowires for quantum computation studies**

Biplab Bhattacharyya,[1,2] V. P. S. Awana,[1,2] T. D. Senguttuvan,[1,2] V. N. Ojha,[1,2] Sudhir Husale[1,2]*

[1] Academy of Scientific and Innovative Research (AcSIR), National Physical Laboratory, Council of Scientific and Industrial Research, Dr. K. S Krishnan Road, New Delhi-110012, India.

[2] National Physical Laboratory, Council of Scientific and Industrial Research, Dr. K. S Krishnan Road, New Delhi-110012, India.

*E-mail: husalesc@nplindia.org

**Contents:**

1. Higher magnification FESEM images

2. HRTEM characterization of W-$Bi_2Se_3$ interface

3. Absence of FIB-induced Halo effect during electrode deposition

4. Magnetoresistance (MR) at 10K for device N5

# 1. Higher magnification FESEM images

Figure S1 shows the higher magnification FESEM images of FIB deposited W electrodes on $Bi_2Se_3$ nanowires. Clean $Bi_2Se_3$ channel lengths can be clearly observed with little contamination or diffusion of W limited very close to the edges of the electrodes within 40 to 50 nm lengths from electrode edge. This is suggestive of the fact that with controlled FIB deposition parameters, electrodes can be deposited at < 500 nm gap lengths to perform such supercurrent studies. We have used optimized and controlled electrode deposition parameters and taken careful steps to keep the Ga ion contamination to minimum during fabrication process. Sharp $Bi_2Se_3$ nanowire sidewalls with missing granular W nano-islands on the substrate surface (in dark black with no white nano-clusters) are clearly visible in these FESEM images. Thus, indicating that most of the TI junction length is free from W contamination.

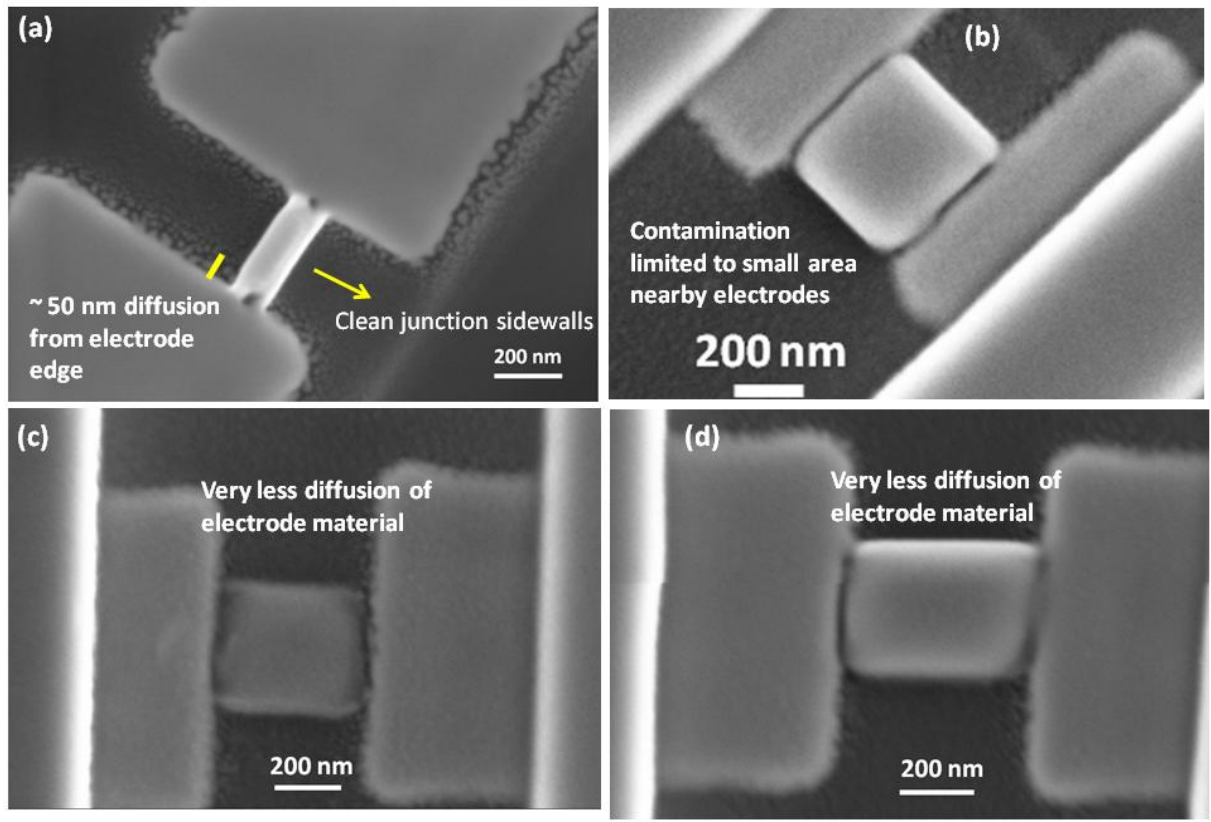

**Fig. S1. Higher magnification FESEM images of $Bi_2Se_3$ nanowire devices. (A)** Device N5 used in manuscript. **(A-D)** Sharp nanowire junction sidewalls are clearly visible in all the fabricated devices. Tiny granular nanoislands are present only at the interface within ~ 50 nm or less, keeping most of the junction length clean.

## 2. HRTEM characterization of W-Bi$_2$Se$_3$ interface

To carry out the HRTEM characterization for interface study of W-Bi$_2$Se$_3$ heterostructure, we deposited flakes of Bi$_2$Se$_3$ on TEM grid and used FIB to deposited W contact on the flake. Here, we show the FESEM images (Fig. S2) of the deposited Bi$_2$Se$_3$ flake and W contact on TEM grid. The high magnification FESEM images in Fig. S2 (B and C) show clear interfaces with no granular islands. The W electrode has very sharp edges and sidewalls, with very restricted diffusion and contamination in the electrodes surroundings. This is a result of substrate dependent deposition optimized conditions.

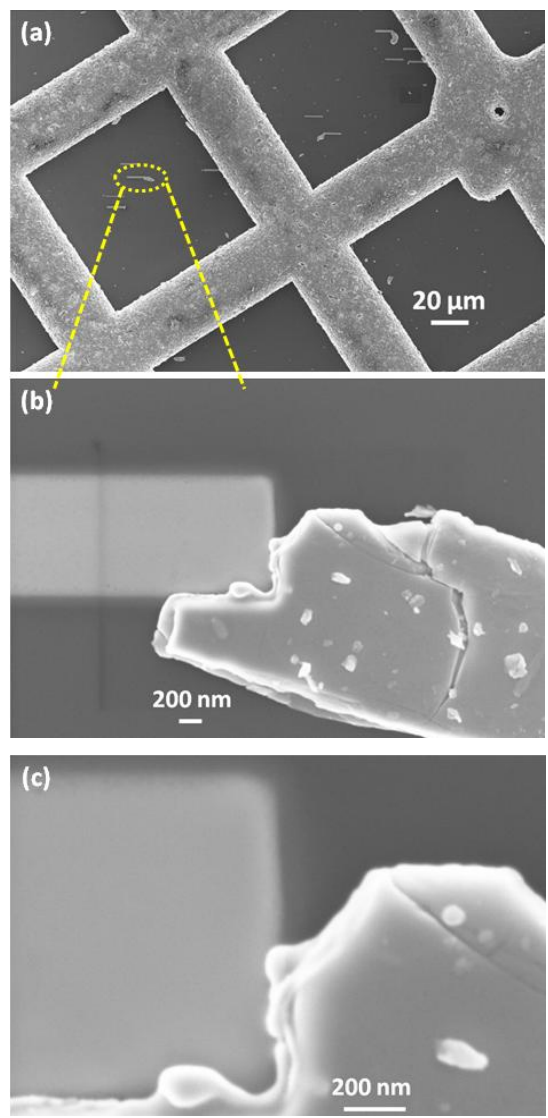

**Fig. S2. FESEM images of Bi$_2$Se$_3$ flake and FIB deposited W contact on TEM grid.** **(A)** Bi$_2$Se$_3$ flake (random shaped) and FIB deposited W electrode (straight horizontal bars) deposited on a TEM grid. **(B)** Higher magnification image of the encircled region in (A) showing the flake and electrode with more clarity. **(C)** Further higher magnification of the interface region between W and Bi$_2$Se$_3$. Very less diffusion or contamination can be observed on the flake or on the surroundings of the electrode. W electrode has sharp edges and sidewalls.

Figure S3 depicts the TEM images of the FIB deposited W electrodes on TEM grid. Figure S2 (a) shows the W electrodes with very good deposition parameters. The magnified electrode and edge region in Fig. S2 (b) shows the small area diffusion of W material in the surroundings of the electrode. The contamination is limited to roughly around 50 nm from the electrode edge.

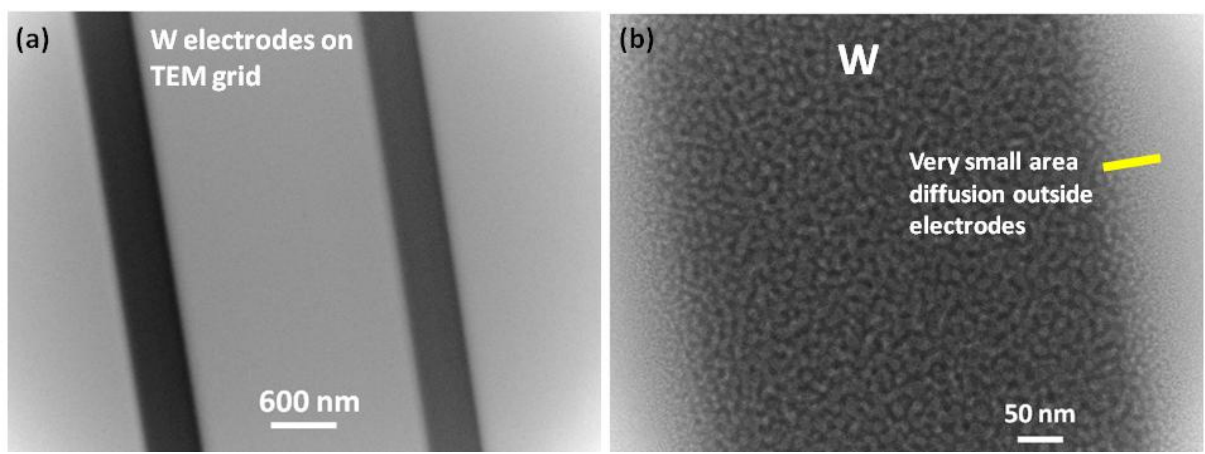

**Fig. S3. TEM images of the W electrodes on TEM grid.** **(A)** W electrodes with very sharp edges and sidewalls. **(B)** Very less diffusion of W material in the neighborhood of electrode. The contamination is limited within ~50 nm from electrode edge.

Figure S4 depicts the HRTEM images of the W electrode in contact with $Bi_2Se_3$ flake deposited on TEM grid. We can clearly see in the magnified view of the interface that crystalline $Bi_2Se_3$ planes start to emerge after some distance from the electrode edge and this distance was again found to be on an average ~50 to 60 nm from the electrode edge. In the area within 50 nm from the electrode edge, no clear crystalline planes were observed, which reflects the fact that the material may have become amorphous in that region and perfect $Bi_2Se_3$ crystal structure may have got distorted in that region. But, overall we find that the contamination is limited to the area closer to the electrode edges.

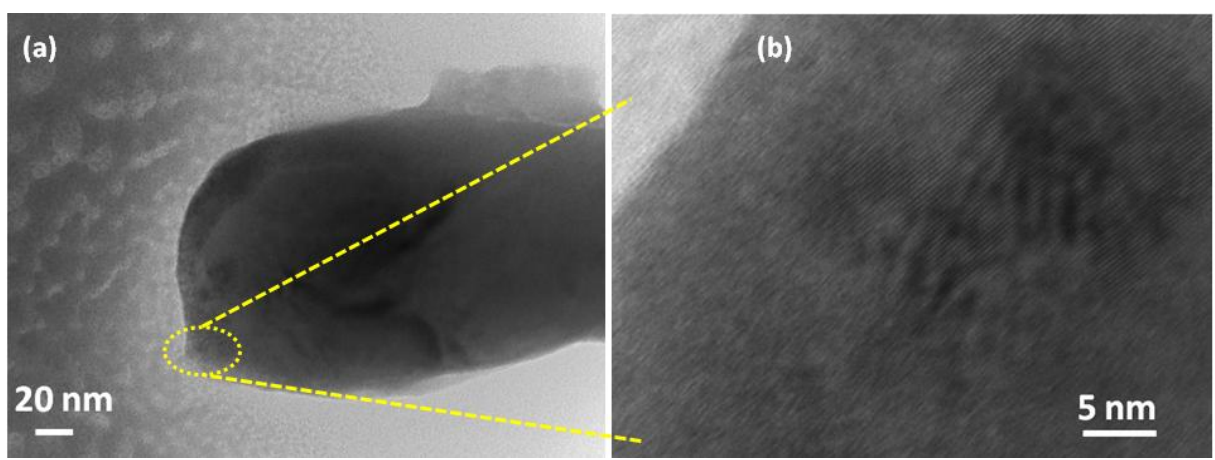

**Fig. S4. HRTEM characterization of the W- Bi2Se3 interface.** **(a)** Interface of W-$Bi_2Se_3$. **(B)** Higher magnification image of the region shown by yellow dashed circle in (A). Clear crystalline planes can be seen to emerge in $Bi_2Se_3$ flake after some distance (~50 to 60 nm) from electrode edge.

## 3. Absence of FIB-induced Halo effect during electrode deposition

The issue of W electrode deposition induced contamination has been taken into consideration in the control experiments. We have deposited only short channel electrodes (with no $Bi_2Se_3$ nanowire) on $SiO_2$/Si substrate (Fig. S5). But, we have found that such nano-gaps are highly insulating in nature, with resistance greater than 50 MΩ. Thus we have concluded that, in case of high quality of nanodevices (very less nano-grains), deposited impurities are insulating and cannot affect the transport properties of TIs. Hence, we believe that the Halo effect is not an issue in our devices, which again strengthens our claim that the observed exciting and unconventional superconductivity results are due to the properties of $Bi_2Se_3$ nanowire junction.

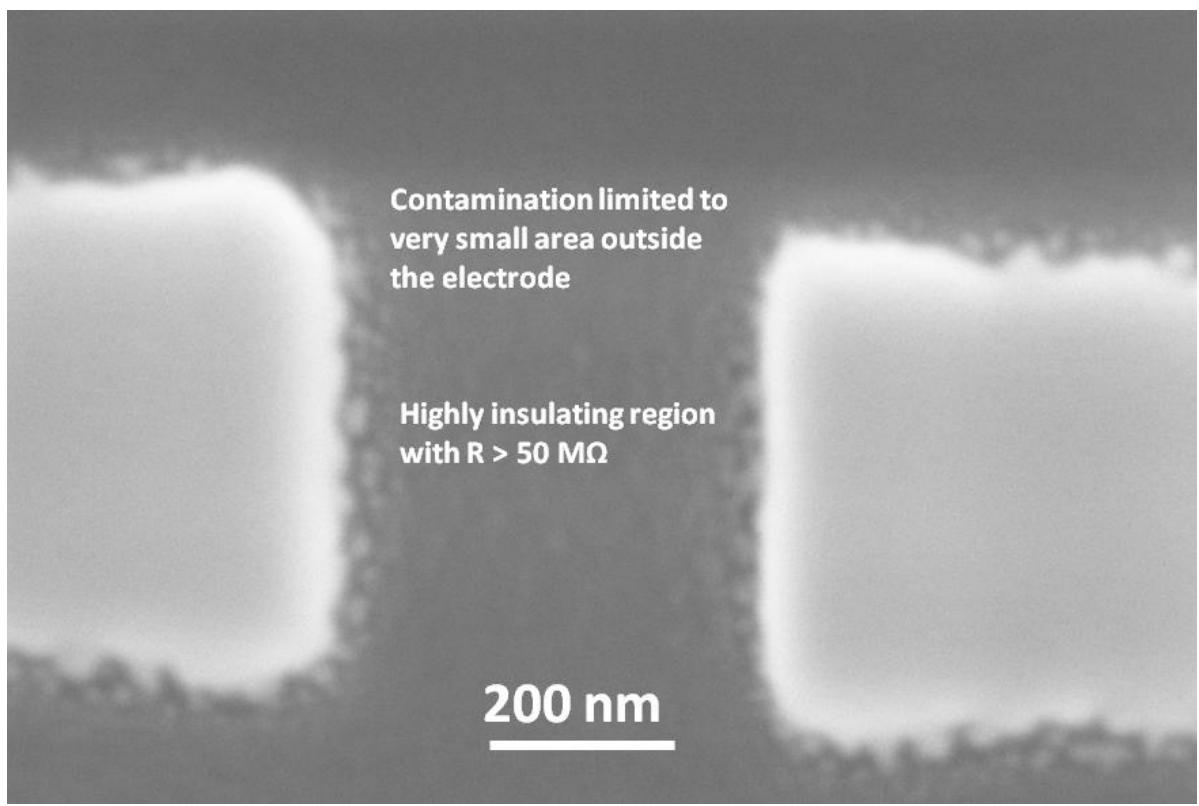

**Fig. S5. Control experiments to confirm the absence of halo effect.** High magnification FESEM image of FIB deposited W electrodes to measure the extent of halo effect through electrical resistance measurement between the nano-gaps. The measured electrical resistance is > 50 MΩ in the nano-gaps.

## 4. Magnetoresistance (MR) at 10K for device N5

We have analysed the MR data for device N5 in Fig. 1B of the main manuscript based on the theory of TIs and their behaviour in perpendicular B-field. Figure S6 shows the comparison in MR (R-R (0 T)) at 2 K and 10 K, where the R (0 T) at 2 K and 10 K is 0 Ω and 566.62 Ω, respectively.

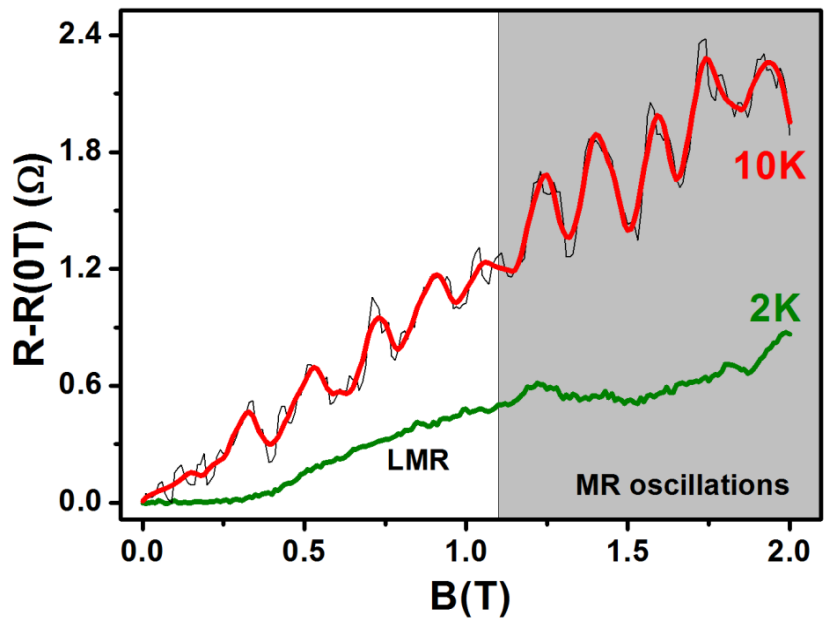

**Fig. S6. MR at 2 K and 10 K for device N5.** Since the nanowire becomes superconducting at 2 K, a zero resistance is observed at lower B-fields with LMR and MR oscillations emerging at higher B-fields. For 10 K, clear MR oscillations under perpendicular B-field with increasing oscillation amplitude with B-field can be observed.

The MR at 10 K has been smoothed (red colour) only to clearly visualize the features in the signal. A positive MR with increasing oscillation amplitude is visible at 10 K (> $T_c$), where sample N5 is completely in the normal state with zero field resistance around 567 Ω. The behaviour at 10 K corresponds to the SdH effect of very low cyclotronic mass fermions ($\omega_c = eB/m_c$) as under perpendicular B-field electrons perform cyclotronic motion that leads to Landau quantization of these orbits with a frequency spacing of $\omega_c$. The dominance of these oscillations at significantly lower B-fields (~ 2 T) suggests the possibility of very low mass fermions present on the surface of TIs. As we shift to a temperature below $T_c$, the MR at 2K shows a completely superconducting state with zero resistance at low B-fields, which is possibly due to the proximity-induced superconductivity in TI. With increasing B-field, first a linear MR (LMR) is observed and then oscillatory features are observed at fields above 1T. Now, we believe that these features (when superconductivity is lost due to B-field) may be arising from the TSS. Thus, we argue in the manuscript that a topological superconducting (TSC) phase has been realized at 2K.